# Efficient Uncertainty Evaluation of Vector Network Analyser Measurements Using Two-Tier Bayesian Analysis and Monte Carlo Method

Min Wang[1,2], Yongjiu Zhao[1]*, Tian Hong Loh[2]* , Qian Xu[1], Yonggang Zhou[1]
[1] Key Laboratory of Radar Imaging and Microwave Photonics, Ministry of Education, Nanjing University of Aeronautics and Astronautics, Nanjing, China, wangmin_win@126.com
[2] Engineering, Materials & Electrical Science Department, National Physical Laboratory, Teddington, United Kingdom, tian.loh@npl.co.uk

***Abstract*—Antennas are a key element in any communication system and vector network analyser (VNA) is popular tool for charactering antenna impedance bandwidth. In this paper, an efficient uncertainty evaluation method is proposed for VNA measurement based on its uncertainty propagation mechanism using Bayesian analysis and Monte Carlo method. The proposed method is generic and can be applied to VNA with arbitrary number of ports. In order to obtain the complete information of measurement uncertainty distribution, a two-tier Bayesian analytic process is carried out. The proposed method contains three steps. In the first step, the posterior distribution of each uncertainty source of VNA calibrations is deduced by the use of prior and current sample information through the first-tier Bayesian analysis. In the second step, the obtained posterior distributions of uncertainty sources are taken into the Monte Carlo simulation of one-port VNA measurement uncertainties. In the last step, the results obtained in the second step are used as the prior distribution of the secondary Bayesian evaluation, then the evaluation results of the measurement uncertainty can be obtained with the means, variances and skewness of the probabilistic distribution. The numerical analysis using an antenna measurement results demonstrate the high-efficiency and reliability of this proposed method.**

***Index Terms*—VNA, measurement, uncertainty, Bayesian analysis, Monte Carlo simulation.**

## I. Introduction

With the extensive application of the vector network analyzer (VNA) in the microwave device and antenna measurements, the uncertainty evaluation method for VNA measurements is one key research focus. To verify the creditability and comparability of a measurement, the uncertainty of the measured result should be evaluated accordingly. The type A evaluation of uncertainty that based on the statistical method is straightforward and considered to be the most reliable solution [1]. However, the accuracy of this method is influenced by the total number of repeated measurements, which results a time-consuming and hard sledding work due to the tedious VNA measurement procedures and operations of the VNAs. In contrast, the type B uncertainty evaluation method for VNA measurements is generally based on the mathematical analysis using, for example, empirical data, manufacturer's specifications, and reference data from handbooks, uncertainty error model, etc. [2]. Because of its efficiency and practicality, a number of type B uncertainty evaluation methods has been proposed, such as the partial derivative method proposed in [2] – [4], the covariance-based uncertainty analysis in [5] – [7], the root-squaring method [8] – [9] (which is widely used in commercial VNA uncertainty assessment systems),etc. However, from post-processing prospective, all these methods are inefficient due to its heavy data processing burden. Also, their accuracy relies on knowledge of prior data and uncertainty model. Furthermore, the type B uncertainty evaluation method results in the lack of the information of the partial degrees of the measurement uncertainty distribution. Therefore it is essential to find a practical way to evaluate the measurement uncertainties of commercial VNAs, which is both efficient and reliable.

Based on small sample size and a prior knowledge of the uncertainty distributions, an efficient and reliable uncertainty evaluation method for VNA measurements is proposed by combining Bayesian statistical methodology with Monte Carlo simulation. This paper presents a detailed procedures for the VNA measurement uncertainty assessment of one-port antenna devices, and the proposed algorithm is explained in section Ⅱ. As an example, an uncertainty analysis for the measurement results of a single antenna is given in section Ⅲ, and good reliability and high efficiency are proven by comparing with practical measurement. The proposed method can also be applied to a multiport VNA measurements.

## II. Methods

### *A. The Uncertainty Sources and Error Model*

The uncertainty of VNA measurement can be evaluated mainly by obtaining the distributions of uncertainty sources. The measurement uncertainties are mainly originated from non-ideal calibration standards and the random errors caused by noise, cable bending and connection repeatability. For example, if one considers the SOL (Short, Open, Load) calibration method, the error model and the flowchart of uncertainty propagation mechanism for one-port VNA measurements are shown in Fig. 1 and Fig. 2, respectively. The systematical errors taken into account include the directivity error, $E_D$, the source mismatch error, $E_S$, and the reflection tracking error, $E_R$. $T_{cc}$ and $R_{cc}$ are respectively the random error caused by cable bending and connector repeatability, and $E_{NRc}$ is the noise floor at the receiver. As shown in Fig. 2, the random errors are global parameters that exist throughout the VNA measurement process.

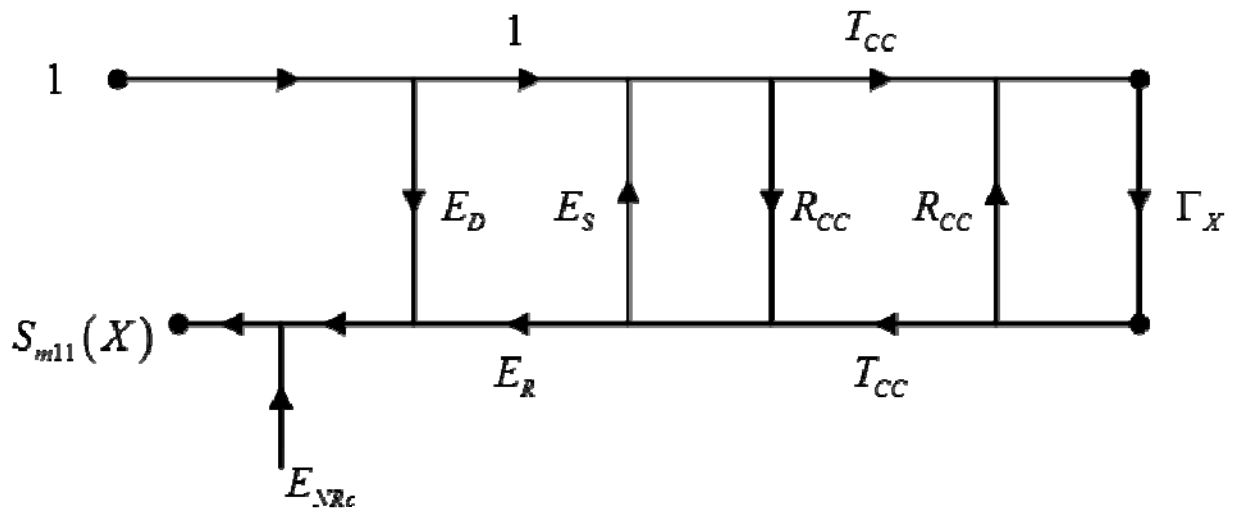

Fig. 1. Uncertainty error model of one-port VNA measurement

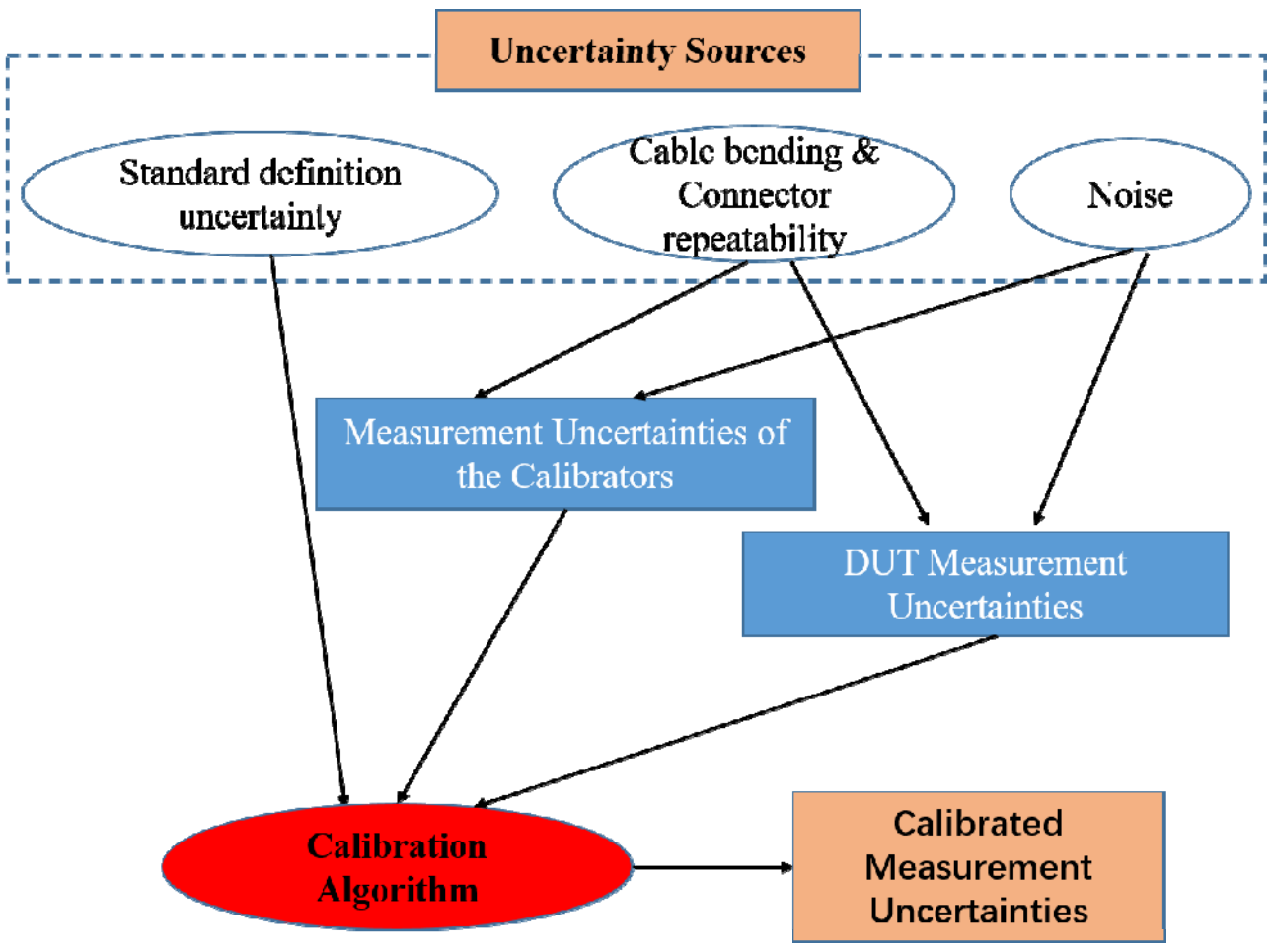

Fig. 2. Flowchart of uncertainty propagation mechanism

## B. *First-Tier Bayesian information fusion*

A critical problem found in VNA measurement uncertainties assessment is that the complexity of error model and the number of uncertainty sources get larger with the port numbers increases. Generally, the main specifications of the commercial VNAs and calibration standards are given in the factory. In practice, the true values may be deviated from the prior data due to ageing loss and environment. Hence, these are not suitable methods for the measurement uncertainty evaluation which inherently has higher request on timeliness. This step is aimed to get more reliable uncertainty information of the calibration kits.

Due to the limitation of the type A or type B uncertainty evaluation, the Bayesian information fusion method is an effective way to get reliable data of uncertainty sources. One notes that the uncertainties of calibration standard definitions could be traced back to the physical parameter uncertainties due to the limitations of hardware manufacturing techniques, connector repeatability and hardware impairment due to wear and tear, e.g. the inaccurate length $l$ of an "Open" standard (see Fig. 3). Furthermore, the uncertainties of $T_{cc}$ and $R_{cc}$ for the error model of one-port VNA measurement could be estimated by repeated measurements with connector repeatability and various cable bending. If one assumes that the physical parameters of standards and the random error items obey normal distribution, the Bayesian analysis process of an arbitrary error source could be described as follows.

The prior distribution function of an error source parameter $X$ can be described as:

$$P(x) = \frac{1}{\sqrt{2\pi}\sigma_0} \exp\left( -\frac{(x - x_0)^2}{2\sigma_0^{\,2}} \right) \tag{1}$$

where $\sigma_0$ and $x_0$ separately respect to the prior empirical mean and variance of $X$.

Consider a recorded variable, $X_{Sample}$ with small sample size of $n$, and its likelihood function is

$$\begin{aligned} L\left(x \middle| x_{Sample}\right) &= \prod_{i=1}^{n} p\left(x_{Sample} \middle| x\right) \\ &= \left( \frac{1}{\sqrt{2\pi} \times std\left(X_{Sample}\right)} \right)^{n} \exp\left( -\frac{\sum_{i=1}^{n} \left(x_{Sample}^{i} - x\right)^2}{2 \times std\left(X_{Sample}\right)^2} \right) \end{aligned} \tag{2}$$

According to Bayesian inference theory, the combined distributing function can be obtained as:

$$\begin{aligned} h\left(x_{Sample}, x\right) &= \frac{1}{\left(\sqrt{2\pi}\right)^{n+1} \sigma_0 \cdot std\left(X_{Sample}\right)^{n}} \cdot \\ & \exp\left( -\frac{1}{2} \left[ \frac{nx^2 - 2nx\overline{X}_{Sample} + \sum_{i=1}^{n} {x_{Sample}^{i}}^{2}}{std\left(x_{Sample}\right)^2} + \frac{x^2 - 2\mu_0 x + \mu_0^{\,2}}{\sigma_0^{\,2}} \right] \right) \end{aligned} \tag{3}$$

Following Bayes' rule, the posterior distribution is proportional to the product of the prior distribution and the likelihood and the posterior distribution function could be obtained as follows,

$$\pi\left(x \middle| x_{Sample}\right) = \left(\frac{A}{2\pi}\right)^{\frac{1}{2}} \exp\left( -\frac{(x - B/A)^2}{2/A} \right) \tag{4}$$

where

$$k_1 = \left[ \left(\sqrt{2\pi}\right)^{n+1} \sigma_0 \cdot std\left(X_{Sample}\right)^{n} \right]^{-1},$$

$$A = n / std\left(X_{Sample}\right)^2 + 1/\sigma_0^{\,2},$$

$$B = \frac{n \cdot \overline{X}_{Sample}}{std\left(X_{Sample}\right)^2} + \frac{\mu_0}{\sigma_0^{\,2}},$$

$$C = \frac{1}{std\left(X_{Sample}\right)^2}\sum_{i=1}^{n} {x_{Sample}^{i}}^2 + \frac{{\mu_0}^2}{{\sigma_0}^2},$$

From Equation (4) one notes that the posterior distribution is normal with mean B/A and variance 1/A, which is in agreement with Equation (5).

$$\mu = \frac{\dfrac{n\cdot \overline{X_{Sample}}}{std\left(X_{Sample}\right)^2} + \dfrac{\mu_0}{{\sigma_0}^2}}{\dfrac{n}{std\left(X_{Sample}\right)^2} + \dfrac{1}{{\sigma_0}^2}},\quad \sigma^2 = \frac{1}{\dfrac{n}{std\left(X_{Sample}\right)^2} + \dfrac{1}{{\sigma_0}^2}} \tag{5}$$

## C. Monte Carlo Simulatin

The VNA measurement and calibration formula of $S_{11}$ for a one-port antenna under test (AUT) are given below:

$$S_{m11}\left(X\right) = E_D + \frac{E_R\left(R_{cc} + \dfrac{{T_{cc}}^2\Gamma_X}{1-R_{cc}\Gamma_X}\right)}{1-E_S\left(R_{cc} + \dfrac{{T_{cc}}^2\Gamma_X}{1-R_{cc}\Gamma_X}\right)} + E_{NRc} \tag{6}$$

$$S_{11_Calibrated} = \frac{S_{m11}-E_D}{E_R + \left(S_{m11}-E_D\right)E_S} \tag{7}$$

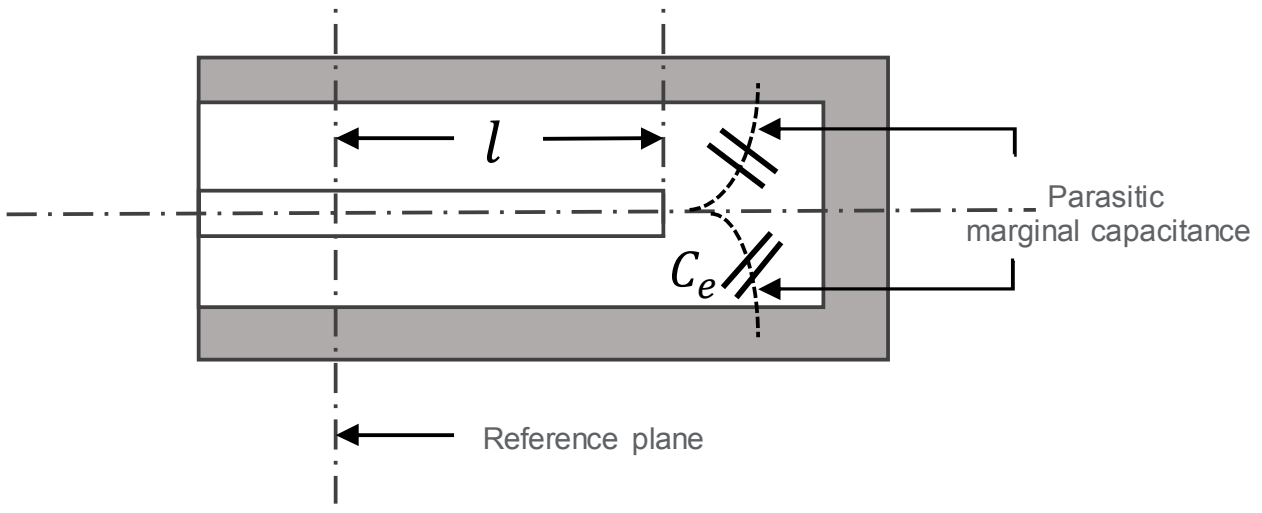

$$C_O = C_0 + C_1 f + C_2 f^2 + C_3 f^3; Z_O = \frac{1}{j2\pi f C_O}$$

$$\Gamma_S = \frac{Z_O - Z_r}{Z_O + Z_r} e^{-j4\pi l_S/\lambda}$$

Fig. 3. Mathematical model of the "Open" standard

According to Figs. 1 and 2, the Monte Carlo simulation of VNA measurements is based on the posterior distribution of the error source calculated by Equations (5) – (7). The mathematical model of calibration standards in the case of standard "Open" calibration kit is shown in Fig. 3, which can be used to obtain the uncertainty information of system error terms through Monte Carlo simulation. With different amplitudes of $S_{11}$ to be measured, different skew degrees of the uncertainty distribution are shown in Fig. 5, where the blue bar shows the probability distribution of simulated results and the red line is the probability distribution curve of normal distribution with simulated mean and standard error. The simulated results will show a more reliable measurement uncertainty distribution which combines the effects of the prior information and the uncertainty source data with small sample size.

## D. Second-Tier Bayesian information fusion

It should be noticed that, from the Monte Carlo simulation results, it can be seen that the traditional way to describe measurement uncertainty distribution only using mean and variance does not suffice due to the lack of distribution information and limited sample numbers. Furthermore, there are still some unknown error resources and complicated propagation mechanisms of measurement uncertainties, so the second-tier Bayesian analysis which contains the evaluation of skewness is introduced to obtain more accurate assessment of measurement uncertainties.

In this step, the Monte Carlo simulated mean $\mu_0'$, variance $\sigma_0'$, skewness coefficient $\lambda_0'$ and its standard error $s$ can be obtained using the commercially available IBM SPSS (Statistical Package for the Social Sciences) [11] and used as the prior uncertainty distribution characteristics of the one-port VNA measurement. Getting AUT measurements with small sample size by repeated calibration and measuring process, the posterior mean and variance can be obtained according to Equation (5), and the further calculation of skewness is described as follows.

As briefly shown in Fig. 4 the skewness of a statistical variable is a measure of the asymmetry of the probability distribution of a real-valued variable about its mean, which can be positive or negative. The definition formula of the skewness is:

$$\gamma_1 = \frac{E\left[\left(X-\mu\right)\right]^3}{\left[E\left(X-\mu\right)^2\right]^{3/2}} = \frac{E\left(X^3\right)-3E\left(X^2\right)E\left(X\right)+2\left[E\left(X\right)\right]^3}{\left[E\left(X^2\right)-\left(E\left(X\right)\right)^2\right]^{3/2}} \tag{8}$$

The skewness of discrete sample data can also be calculated by

$$\gamma_n = \frac{\sqrt{n}\sum_{i=1}^{n}\left(x_{Sample}^{i} - \overline{X_{Sample}}\right)^3}{\left[\sum_{i=1}^{n}\left(x_{Sample}^{i} - \overline{X_{Sample}}\right)^2\right]^{3/2}} \tag{9}$$

A skewed distribution model is developed based on the prior probability distribution characteristics of DUT measurement uncertainties.

$$\pi\left(x,\theta\right) \sim S\left(\mu_0', \sigma_0', \theta\right) \tag{10}$$

where $\theta$ is the undetermined skew coefficient of measurement uncertainty distribution which depends on the prior information, $\lambda$ and its standard error, $s$.

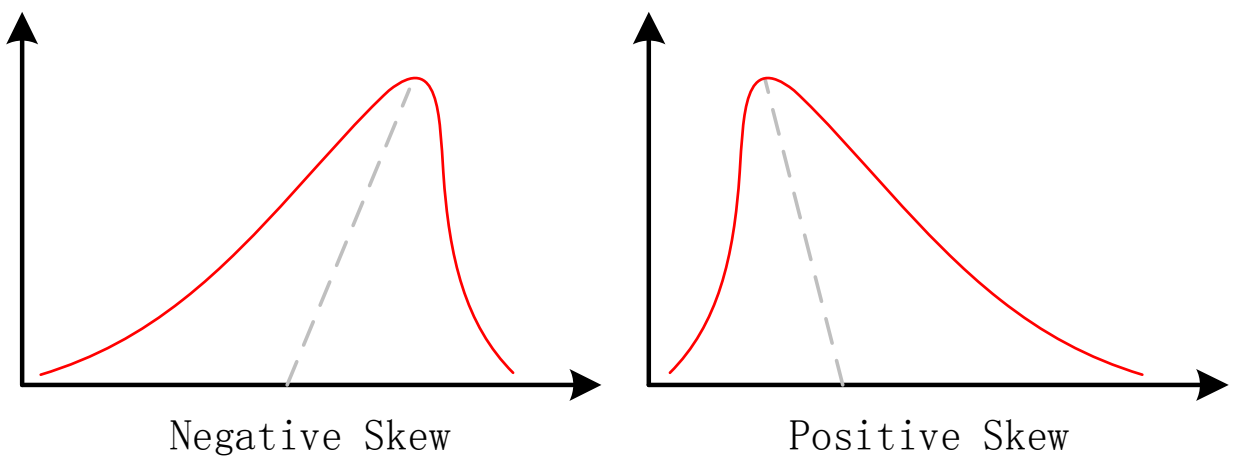

Fig. 4. Schematic diagram of skewed distribution

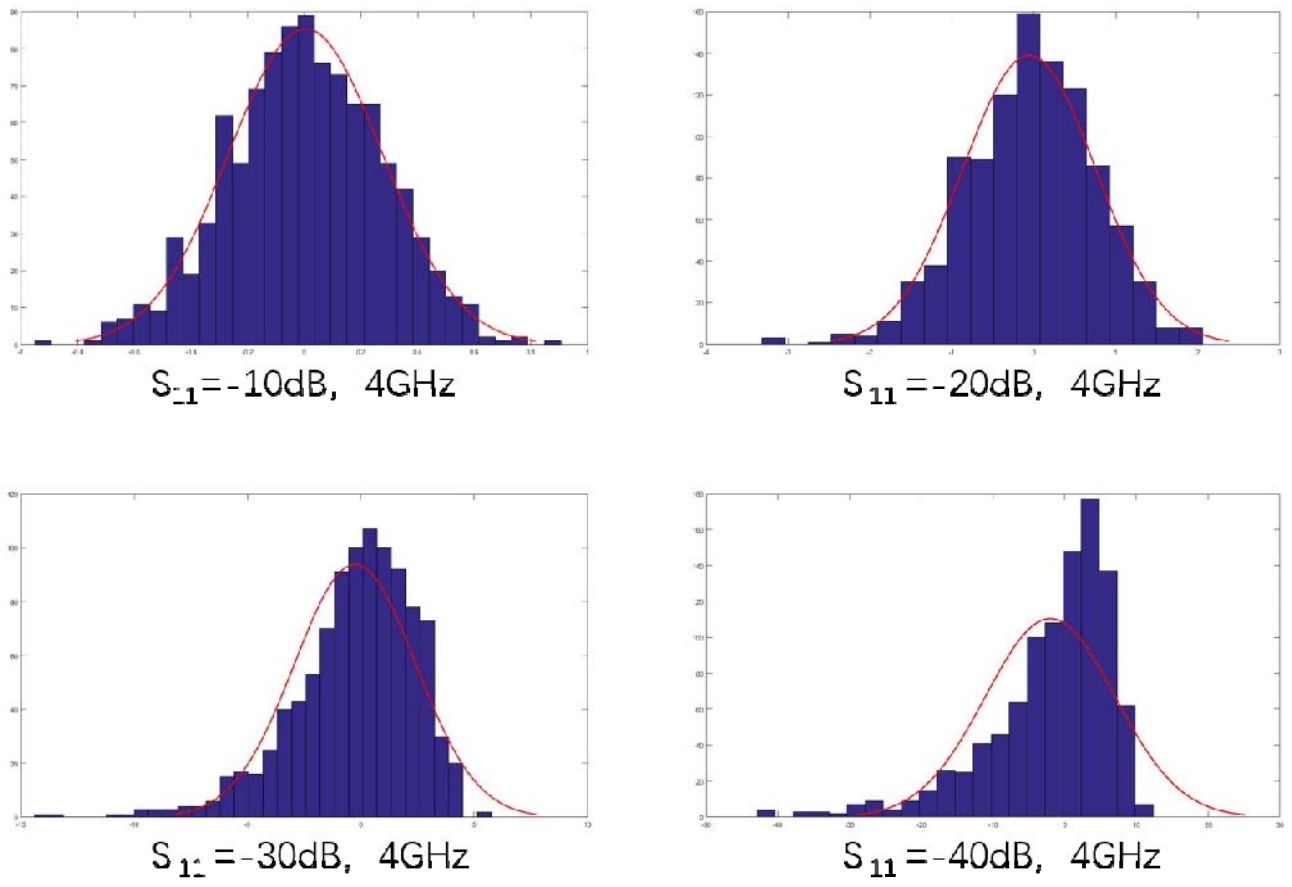

Fig. 5. The frequency histograms of measurement uncertainty distribution of Monte Carlo simulation (Sample Size = 1000)

For simplicity, one assumes that the sample size of the Monte Carlo simulation is large enough that $\theta$ becomes a constant, and the posterior, hence $\hat{\gamma}(\theta)$ can be eventually calculated using the linear Bayesian analysis based on the assess reliability theory [12],

$$\hat{\gamma}(\theta)=\frac{\hat{\mu}_3(\theta)-3\hat{\mu}_2(\theta)\hat{\mu}_1(\theta)+2(\hat{\mu}_1(\theta))^3}{\left[\hat{\mu}_2(\theta)-(\hat{\mu}_1(\theta))^2\right]^{\frac{3}{2}}} \tag{11}$$

where,

$$\hat{\mu}_k(\theta)=\beta_k\overline{X_{Sample}}^{\,k}+(1-\beta_k)\mu_k,\ k=1,2,3.$$

$$\beta_k=\frac{n\left(\tau_k^{\,2}-\mu_k^{\,2}\right)}{n\left(\tau_k^{\,2}-\mu_k^{\,2}\right)+\mu_{2k}-\tau_k^{\,2}}$$

$$\tau_k^{\,2}=E\left[\left(E\left(X^k\,|\theta\right)\right)^2\right]=\mu_k^{\,2}+D\left[E\left(X^k\,|\theta\right)\right]$$

$$\mu_k=E\left(X^k\right)$$

## III. Experimental Verification

To demonstrate the efficiency of the proposed method, some numerical analysis examples are given in this section. As shown in Fig. 6, under the assumption that the $S_{11}$ of AUT is -30 dB over the frequency range between 1 and 4 GHz, it can be seen that the accuracy of uncertainty evaluation results has been greatly improved for an uncertainty source characteristics with small sample size. Meanwhile, the prior skewness information can also be obtained from simulated results. Obviously, the information of skewness is an essential characteristic to describe a non-normal distributed statistic accurately, as shown in Fig. 5. To verify the whole proposed method, a practical work to the $S_{11}$ measurement uncertainties evaluation of a single antenna. The small sample size we separately take in first-tier and second-tier Bayesian analysis is 50, which is easy to operate achieve. In addition, a high accuracy of the evaluated results is shown in Fig. 7 to demonstrate the validity and efficiency of the proposed approach. It is worth noting that due to the inheritability of the Bayesian analysis, the proposed method would greatly improve the efficiency of the uncertainty evaluation processes that a metrology engineer may wish to perform.

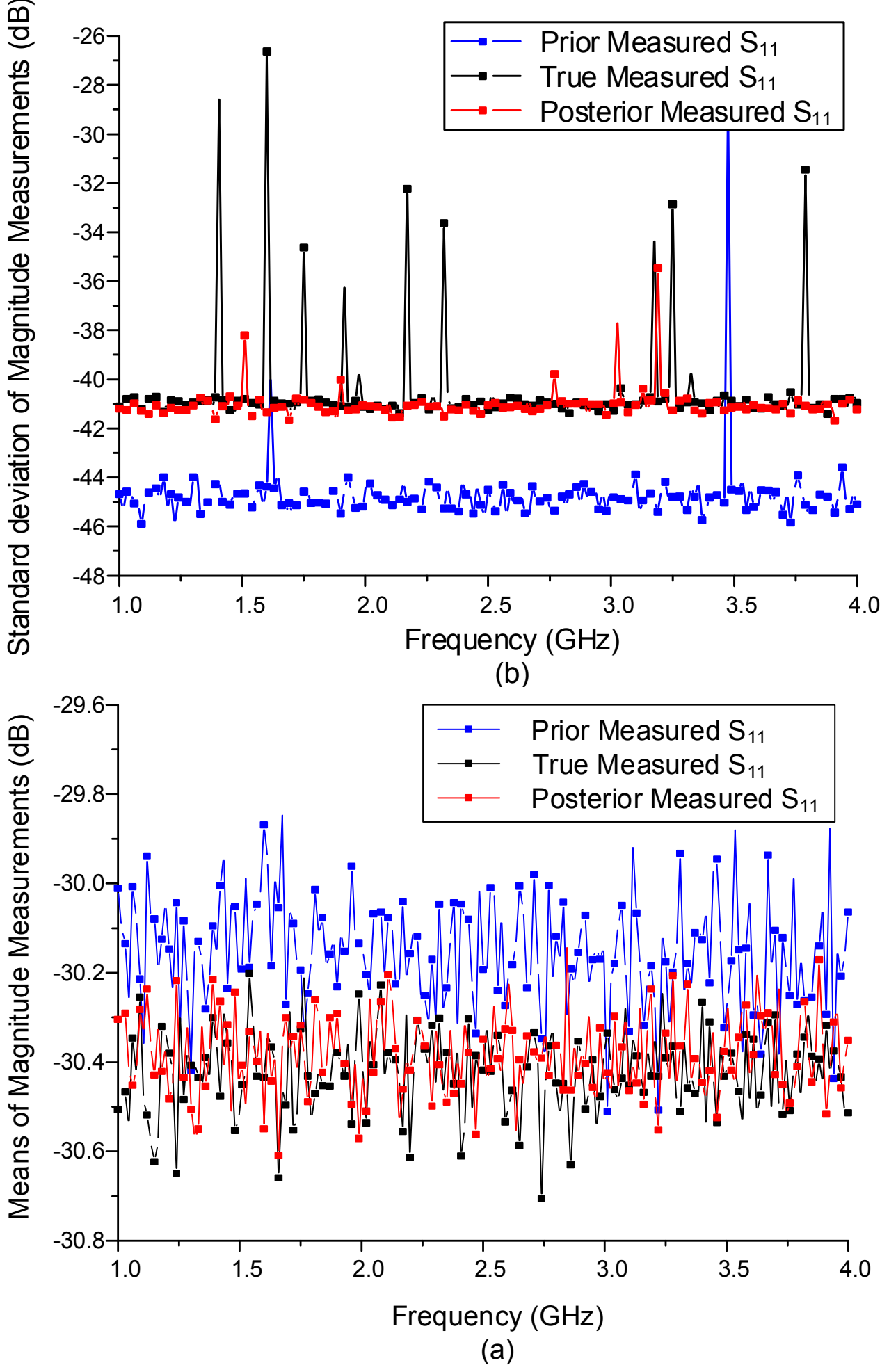

Fig. 6. Comparison of uncertainty distribution characteristics after the first-tier Bayesian information fusion and Monte Carlo simulation (Sample Size = 50)

## IV. Conclusion

In this paper, an efficient and reliable two-tier Bayesian analysis method combined with Monte Carlo simulation for VNA measurement uncertainty evaluation is proposed. By considering a small sample size of error resource characteristics and repeated AUT $S_{11}$ measurement, the evaluation accuracy of measurement uncertainty distribution is significantly improved. This has been demonstrated by numerical analysis using practical antenna measurement. Due to the high-efficiency and inheritability of the Bayesian analysis, this method can be used in both commercial VNA measurement perform evaluation and reliable VNA measurement uncertainties assessment in metrology laboratory.

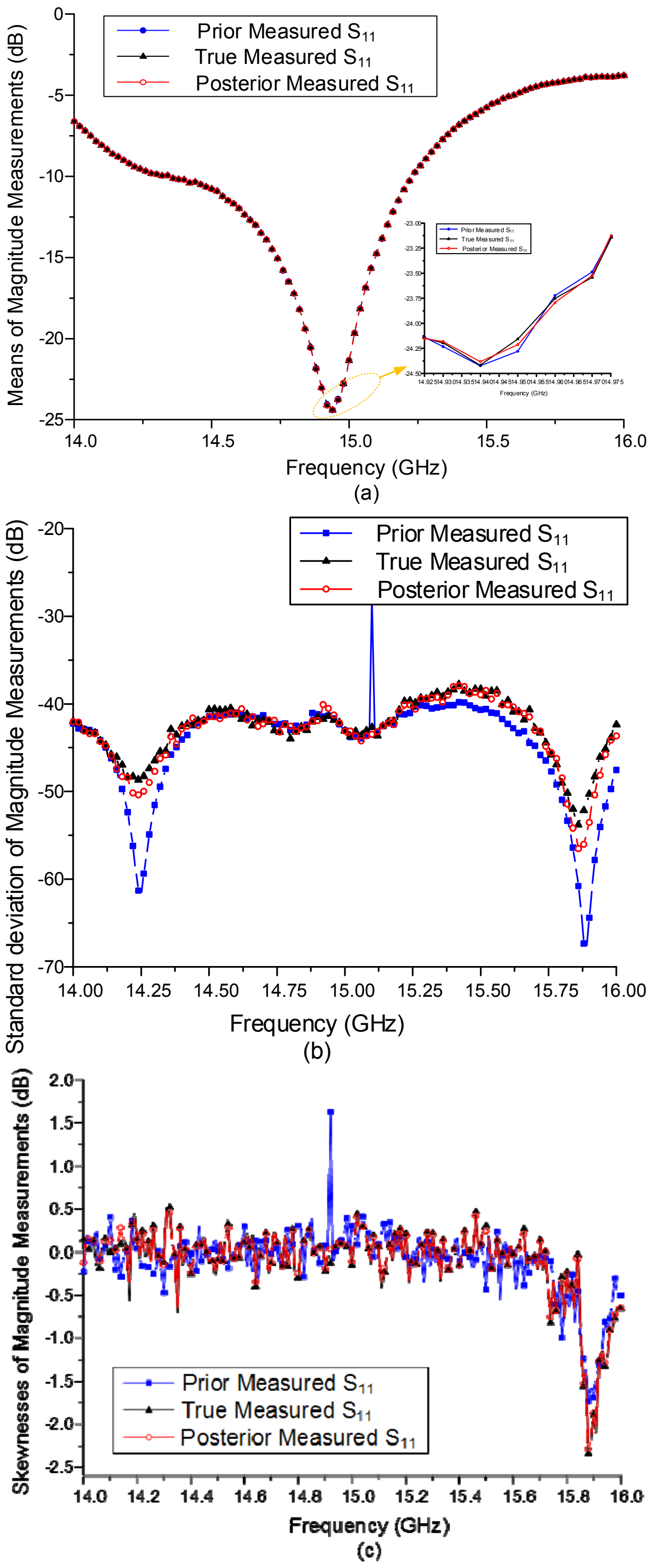

Fig. 7. Comparison of uncertainty distribution characteristics after the second-tier Bayesian information fusion (Sample Size = 50)

## Acknowledgment

This work was supported by the National Science Foundation of China (under project number 61471193), the China Scholarship Council and the Funding of Jiangsu Innovation Program for Graduate Education (under project number KYLX15_0284). The work of T H Loh was supported by the 2017 - 2020 National Measurement System Programme of the UK government's Department for Business, Energy and Industrial Strategy (BEIS), under Science Theme Reference EMT17 of that Programme.